\documentclass[12pt]{article}
\usepackage{times}
\usepackage{geometry}
\usepackage[utf8]{inputenc}
\usepackage{enumitem,amssymb}
\usepackage{ragged2e}
\newlist{thematic}{itemize}{8}
\setlist[thematic]{label=$\square$}
\usepackage{pifont}

\usepackage{graphicx}
\usepackage{cite}
\usepackage[pdftex]{hyperref}
\usepackage{titlesec}
\titlespacing*{\section}{0pt}{0.5\baselineskip}{0.5\baselineskip}

\newcommand{\Xmax}{\ensuremath{X_\mathrm{max}}}

\begin{document}
\raggedright
\huge
Astro2020 Science White Paper \linebreak

High-Energy Galactic Cosmic Rays \linebreak
\normalsize

\noindent \textbf{Thematic Areas:} \hspace*{60pt} $\square$ Planetary Systems \hspace*{10pt} $\square$ Star and Planet Formation \hspace*{20pt}\linebreak
$\square$ Formation and Evolution of Compact Objects \hspace*{31pt} $\boxtimes$ Cosmology and Fundamental Physics \linebreak
  $\square$  Stars and Stellar Evolution \hspace*{1pt} $\square$ Resolved Stellar Populations and their Environments \hspace*{40pt} \linebreak
  $\square$    Galaxy Evolution   \hspace*{45pt} $\boxtimes$             Multi-Messenger Astronomy and Astrophysics \hspace*{65pt} \linebreak
  
\textbf{Principal Author:}

Name: Frank G. Schroeder
 \linebreak						
Institution: Bartol Research Institute, Department of Physics and Astronomy, University of Delaware, Newark, DE, USA
 \linebreak
Email: fgs@udel.edu
 \linebreak
Phone: +1 302 831 1521
 \linebreak
 
\textbf{Co-authors:}\\
T.~AbuZayyad$^1$, L.~A.~Anchordoqui$^2$, K.~Andeen$^3$, X.~Bai $^4$, S.~BenZvi$^5$, D.R.~Bergman$^1$, A.~Coleman$^6$, H.~Dembinski$^7$, M.~DuVernois$^8$, T.~K.~Gaisser$^6$, F.~Halzen$^8$, A.~Haungs$^9$, J.~L.~Kelley $^8$, H.~Kolanoski$^{10}$, F.~McNally$^{11}$, M.~Roth$^9$, F.~Sarazin$^{12}$, D.~Seckel$^6$, R.~Smida$^{13}$, D.~Soldin$^6$, D.~Tosi$^8$
 \linebreak
 
$^1$ University of Utah, $^2$ City University of New York, $^3$ Marquette University, $^4$ South Dakota School of Mines and Technology, $^5$ University of Rochester, $^6$ University of Delaware, $^7$ Max-Planck-Institute for Nuclear Physics, Heidelberg, $^8$ Wisconsin IceCube Particle Astrophysics Center, $^9$ Karlsruhe Institute of Technology,$^{10}$ Humboldt University $^{11}$ Mercer University, $^{12}$ Colorado School of Mines, $^{13}$ University of Chicago 
 \linebreak
 
\textbf{Abstract:}\\
The origin of the highest energy Galactic cosmic rays is still not understood, nor is the transition to EeV extragalactic particles. 
Scientific progress requires enhancements of existing air-shower arrays, such as: 
IceCube with its surface detector IceTop, and the low-energy extensions of both the Telescope Array and the Pierre Auger Observatory.\\

\pagebreak

\justifying
\section{Overview and Science Goals}\vskip-2mm
The most powerful accelerators of cosmic rays in our Milky Way have not yet been revealed. 
In the past decade, several experiments have found evidence that the highest energy cosmic rays (CRs) -- above a few EeV -- seem to be of extragalactic origin \cite{Abbasi:2016kgr, Aab:2018mmi}. 
However, there are major open questions regarding Galactic cosmic rays (GCRs) in the energy range from $\sim$0.1~PeV to $\sim$1~EeV ($10^{14}$--$10^{18}$~eV): 
the source of the highest-energy GCRs remains unidentified; the maximum energies of various acceleration mechanisms are uncertain; the Galactic-extragalactic transition and several features in the CR energy spectrum, such as the Knee, the Second Knee, and the Ankle, are not well understood. 
These questions can be addressed through improved GCR measurements in conjunction with gamma-ray and neutrino observations. 
Bringing multi-messenger astrophysics to maturity requires a consistent understanding and sufficient measurement accuracy for all messengers. 
In particular, the mass composition of CRs is highly sensitive to their origin and propagation \cite{KampertUnger2012}, and the discovery of mass-dependent anisotropies may reveal the most energetic sources in the Galaxy, either directly or in combination with other messengers.
In the next decade, detector enhancements will bring essential improvements in accuracy for GCRs, and may add a GCR channel to neutrino and gamma experiments, transforming them to multi-messenger detectors.

\begin{figure}[t]
    \centering
    \includegraphics[width=0.932\textwidth]{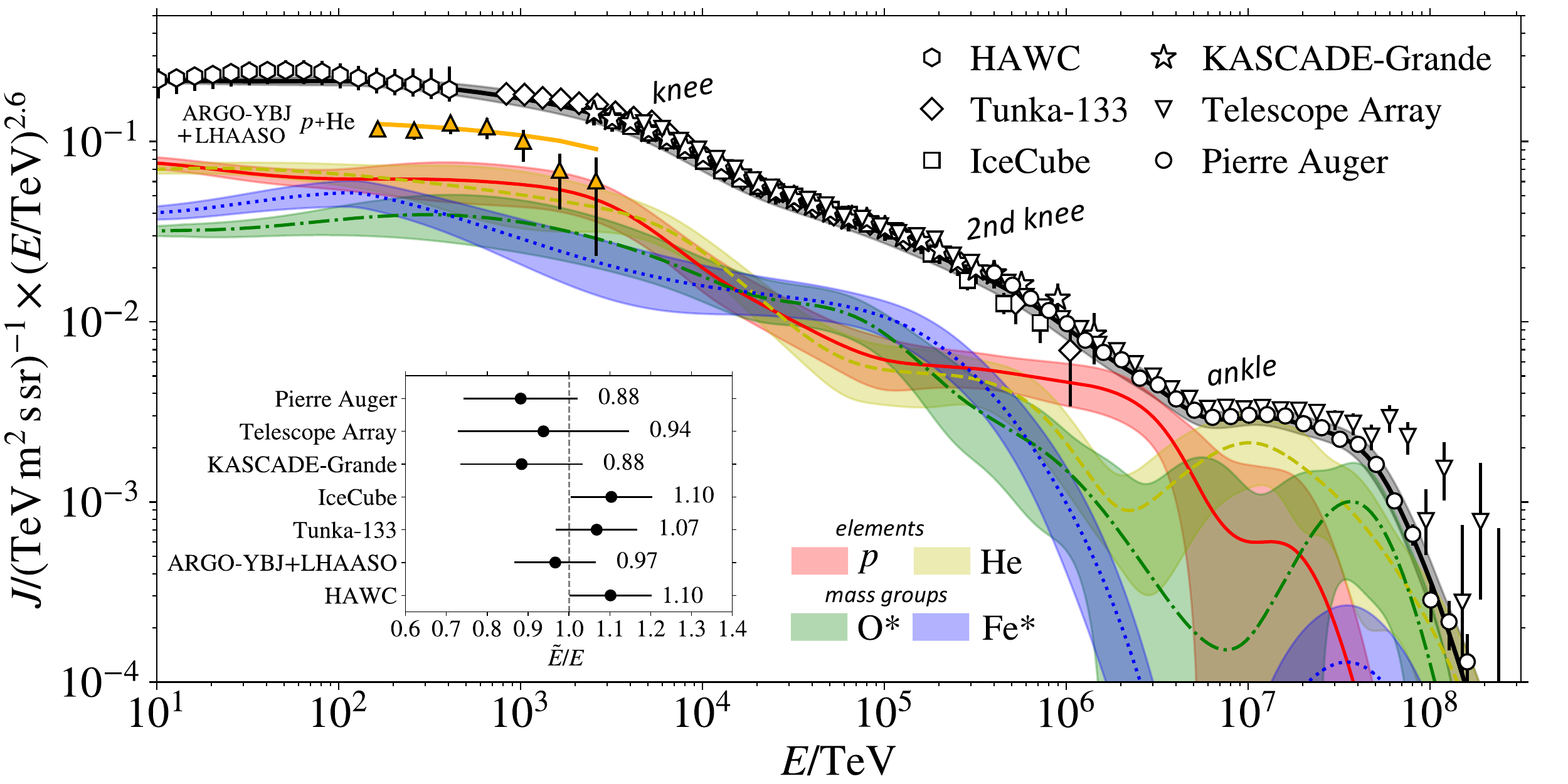}
    \vskip-3mm
    \caption{Cosmic-ray energy spectrum and mass composition.
    This plot reflects a recent attempt for a combined fit of flux and composition measurements by different experiments, where the individual spectra \cite{Bartoli:2015vca,Alfaro:2017cwx,Prosin:2014dxa,Plum:2018,Schoo:2015oxd,IvanovICRC2015,Fenu:2017hlc} have been multiplied by a constant (inset) to adjust them to a common energy scale.
    The bands are one-sigma uncertainties derived from published experimental data (see \cite{GSF_ICRC:2017} for details). 
    Features such as the Knee, Second Knee, and Ankle mark softenings or hardenings of the spectrum which can be approximated by a power law in between these features.
    For future progress in GCR science, it is important to reduce the uncertainties by improving hadronic interaction models and enhancing air-shower arrays to perform hybrid measurements.}\vskip-1mm
    \label{fig_spectrum}
\end{figure}

\begin{figure}[t]
    \centering
    \includegraphics[width=\textwidth]{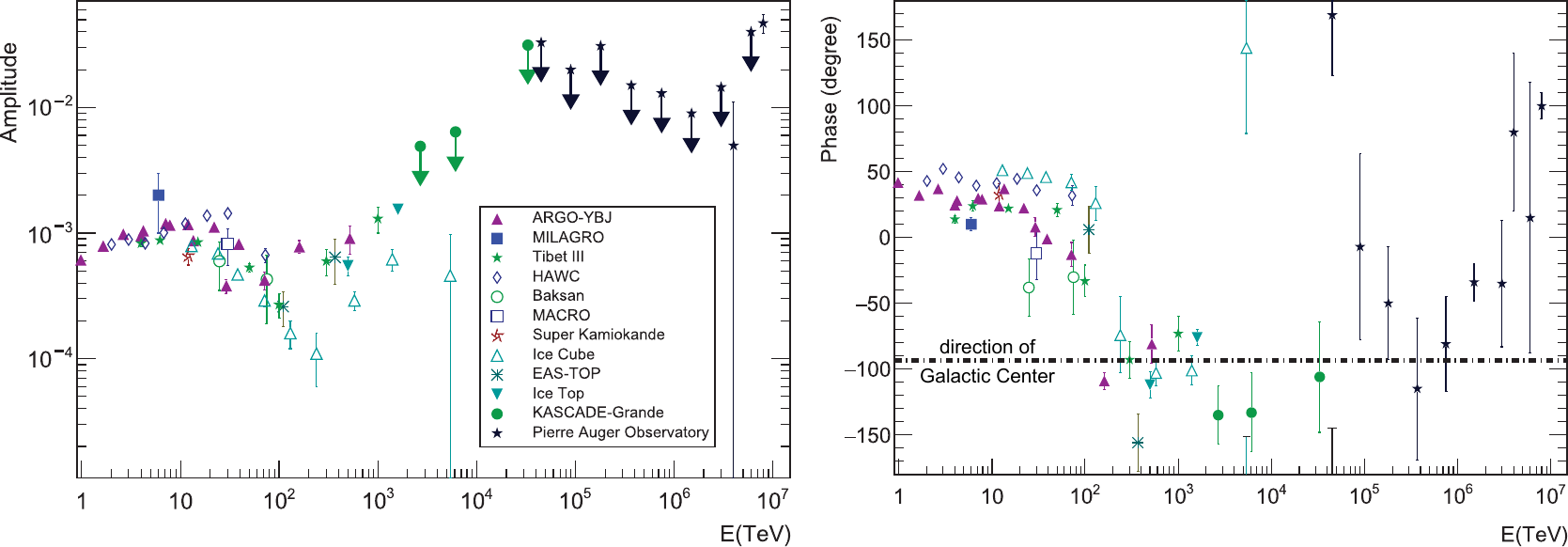}
    \vskip-2mm
    \caption{Large-scale cosmic-ray anisotropy: amplitude (left) and phase (right) of the first harmonic measured by various air-shower experiments, each covering only a part of the sky. 
    In the amplitude plot, upper limits are shown for measurements with a significance of less than three sigmas. 
    The first phase flip may indicate that GCR above 100~TeV belong to at least one different population. 
    The second phase flip at a few EeV may be a signature of the transition from Galactic to extragalactic sources. 
    Figure adopted from \cite{Apel:2019afz}, data from \cite{Apel:2019afz, Ambrosio:2002db, Amenomori:2005dy, Amenomori:2017jbv, Guillian:2005wp, Abdo:2008aw, Aglietta:2009mu, Alekseenko:2009ew, IceCube:2010aug, IceCube:2012feb, IceCube:2013mar, Bartoli:2015ysa, Bartoli:2018ach, Aab:2017tyv, hawclsa2017}.}\vskip-2mm
    \label{fig_anisotropy}
\end{figure}

In this white paper, opportunities for research regarding high-energy GCRs will be presented and discussed. 
The most important goals are as follows:
\begin{itemize}[itemsep=4pt,topsep=4pt]
    \item Determine the maximum acceleration energy for GCRs and understand the transition from Galactic to extragalactic cosmic rays.
    This may be accomplished by discovering the most energetic source in the Milky Way.
    The Galactic Center is a promising candidate \cite{Aharonian:2006au, Abramowski:2016mir}, which can be tested by combining mass-resolved GCR measurements with searches for PeV neutrinos and photons.
    Alternative hypotheses include re-acceleration scenarios and the high-energy tail of a smooth source distribution, e.g., supernova shock fronts. 
    \item Clarify the nature of features in the energy spectrum (Fig.~\ref{fig_spectrum}): 
    Is the Knee related to different populations of sources or magnetic diffusion of the CRs during propagation? 
    Is the Second Knee a feature of the Galactic-to-extragalactic transition, and is it related to the maximum acceleration energy of a particular source population or to the propagation of GCRs? 
    How are features in the energy spectrum linked to features in the large-scale anisotropy (Fig.~\ref{fig_anisotropy})?
    \item Describe the full picture of sources and propagation of GCRs using joint interpretations with other astrophysical messengers and astronomical observations, such as measurements of Galactic magnetic fields. 
    A Galactic contribution to the astrophysical neutrino flux can provide a lower limit on the maximum acceleration energy~\cite{Anchordoqui:2013qsi}.
    \item Resolve apparent differences between CR measurements by various air-shower arrays.     
    What is the reason for differences in their energy scales \cite{GSF_ICRC:2017}? 
    Why is the knee-like feature at $0.7$~PeV observed only by ARGO \cite{Bartoli:2015vca}, and how is it related to the Knee at 3~PeV? 
    Are there differences in GCRs from different parts of the sky further than the small difference in the absolute flux due to the large-scale anisotropy? 
    \item Use GCRs to access particle physics beyond the phase space of current human-made accelerators. A better understanding of hadronic interactions is a science goal by itself, and is also required to fix the mismatch in muons between simulations and data \cite{Dembinski:2019uta, Gonzalez:2018, DeRidder:2017alk, Soldin:2018vak, Aartsen:2015nss}, which constitute a major systematic uncertainty for the interpretation of GCR measurements \cite{Aab:2016hkv}.
    
\end{itemize}


\section{State of the Art in High-Energy GCR Science}\vskip-2mm
To understand how these science goals can be targeted by more accurate mass measurements of GCRs and by multi-messenger astrophysics, let us first review the recent progress in CR science.

GCRs have a strongly mixed composition of elements from proton to iron at all energies (Fig.~\ref{fig_spectrum}; see \cite{Gaisser:2016uoy,Mollerach:2017idb, Bluemer:2009zf, Haungs:2003jv, Spurio:2018knn, Hofmann:2018zma} for reviews). 
The exact composition varies with kinetic energy. 
During their propagation, GCRs diffuse in Galactic magnetic fields depending on their rigidity, which translates into a mass-dependence, since GCRs are fully ionized nuclei. 
The magnetic deflection leads to an almost isotropic arrival direction at Earth (Fig.~\ref{fig_anisotropy}).
Weak anisotropies have been measured for the all-particle flux \cite{Deligny:2018blo, Aartsen:2018ppz, Ahlers:2016rox, Apel:2019afz}, but are difficult to interpret because the present accuracy of air-shower arrays does not allow for efficient per-event classification of the mass of GCRs.

Features in the energy spectrum  (Fig.~\ref{fig_spectrum}) and large-scale anisotropy (Fig.~\ref{fig_anisotropy}) mark the 100~TeV to 1~EeV range of high-energy GCRs as distinct. 
Above 100~TeV, the phase of the first harmonic of the large-scale anisotropy flips towards the direction of the Galactic Center. 
Structures in the heavier components suggest that GCR at 100~TeV and at a few PeV might belong to different populations of sources. 
However, we do not yet know why the Knee, as the most prominent feature in the energy spectrum, does not coincide with a significant feature in the anisotropy measurements. 
Current anisotropy searches are not yet sensitive enough in this energy range \cite{Aartsen:2016ivj}.

Around 100~PeV the heavy component in the energy spectrum becomes softer \cite{KGheavyKnee2011}, and at about the same energy the light component becomes harder \cite{2013ApelKG_LightAnkle}. 
The relation to the Second Knee in the all-particle spectrum detected by several experiments at two to three times higher energy needs further investigation \cite{Abbasi:2018xsn, Prosin:2014dxa}.
Are these separate features or one feature extending over a larger energy range? 
Moreover, a better comparison of the absolute energy scales of different experiments is needed to check for their compatibility.  
A detailed understanding is of particular importance since this energy range around 100~PeV could mark the start of the Galactic-to-extragalactic transition. 

The change of the large-scale anisotropy around $1\,$EeV is likely related to the Galactic-to-extragalactic transition, too \cite{Abreu:2012ybu}. 
Recent observations indicate that this transition seems to be completed at an energy below the Ankle.
At higher energies, the large-scale dipole anisotropy points away from the Galactic Center.
At a few EeV, no strong anisotropy is observed, although anisotropy is expected if the sources are mostly Galactic in the EeV range \cite{Abbasi:2016kgr, Aab:2018mmi}. 
To distinguish and resolve several models suggesting a lower transition energy, e.g., \cite{Abu-Zayyad:2018btv, Mollerach:2018lkt,Thoudam:2016syr,Globus:2015xga,Anchordoqui:2018qom}, more accurate experimental data is required.
The transition energy from Galactic to extragalactic cosmic rays is of interest by itself, since it may either be related to the propagation of GCRs or mark their maximum acceleration energy.
Since the most energetic GCRs are expected to consist of heavier nuclei, a per-event mass classification is vital to separate them from the extragalactic CRs, which consist mostly of light nuclei at energies below the Ankle~\cite{Aab:2016htd}.

Theoretical models and observations suggest that GCRs are predominantly accelerated by supernova shock fronts, but it is difficult to explain the origin of the most energetic GCRs with this mechanism \cite{Blasi:2013rva}. 
The Galactic Center might be a more powerful accelerator than supernovae, as gamma-ray observations suggest a maximum energy of at least $1\,$PeV for protons \cite{Aharonian:2006au, Abramowski:2016mir}.
Typically it is assumed that any electromagnetic acceleration process will have a maximum energy for nuclei of $Z$ times the maximum for protons, where $Z$ is the charge number of the CR nucleus.  
To account for the most energetic GCRs in the Milky Way, a maximum energy of $Z$ times at least a few PeV if not several 10s of PeV may be required, depending on the transition energy to extragalactic CRs, and the mass composition of the GCRs at the transition energy. 
Measuring the spectra of individual mass groups may resolve these so-called Peters cycles \cite{PetersCycles1961} of the $Z$ dependence of the maximum acceleration energy, and indicate how many source populations exist. 

In summary, we do not yet know what accelerates the most energetic GCRs and need to understand the transition to extragalactic CRs. 
In addition to multi-messenger observations, the key for progress is to increase the measurement accuracy for GCRs by enhancing detectors for air showers.\vskip-40mm

\section{Progress and Challenges in Detection Techniques}\vskip-2mm
The measurement of CRs at energies of a PeV and beyond remains challenging because only indirect measurements by air showers can acquire sufficient statistics. 
The interpretation of these air showers is subject to systematic uncertainties that depend on the type and coverage of the used instrumentation.
Deficits of hadronic interaction models \cite{Dembinski:2019uta} constitute a major systematic uncertainty. 
Improving the models requires input by accelerator experiments and by enhanced air-shower arrays providing hybrid measurements of all accessible shower components. 
According to the paradigm of shower universality \cite{Lipari:2008td, Lafebre:2009en}, air showers can be described in good approximation by only a few parameters, such as their direction, energy, muon content and the depth of the shower maximum, \Xmax.
Shower-to-shower fluctuations and the similarity of showers initiated by different primary nuclei lead to intrinsic systematic uncertainties.
Therefore, when using state-of-the-art techniques, maximizing the accuracy for the properties of the primary particle requires the simultaneous measurement of at least the energy, the number of muons, and \Xmax. 

In the past, the accuracy of most air-shower arrays for GCRs was limited by using a single technique, such as particle or air-Cherenkov detectors. 
For example, KASCADE \cite{Apel:2013uni} and IceTop \cite{IC40_Composition} statistically estimated the mass composition based on the muon content.
Tunka, TALE, and  Auger used \Xmax\ for this purpose \cite{Prosin:2014dxa, Abbasi:2018nun, Abreu:2013env}. 
While mass measurements based on the muon content have a high precision, they suffer most from uncertainties due to deficiencies in the hadronic models \cite{Aab:2016hkv}. 
Even though the LHC reaches the center-of-mass energy of GCR collisions with the atmosphere, its experiments do not cover the complete phase space relevant for air showers \cite{Ulrich:2010rg, Parsons:2011ad, dEnterria:2011twh, Ostapchenko:2014mna}.
The interpretation of \Xmax\ is presently more accurate since the electromagnetic component of air showers is better understood \cite{Aab:2016zth}. 
For EeV energies, the fluorescence technique yields the best accuracy for determining the energy and \Xmax\ \cite{KampertUnger2012, Aab:2017ctu, Abbasi:2018nun}, and below about a 100~PeV air-Cherenkov detectors are used for \Xmax\ measurements \cite{Abbasi:2018xsn, TAIGA_2014}.
Both methods require clear, moonless nights. 
Recently there has been progress in the radio technique providing a 24/7 instrumentation for \Xmax\ around 100~PeV \cite{2012ApelLOPES_MTD, 2014ApelLOPES_MassComposition, LOFARNature2016, TunkaRex_Xmax2016, HuegeReview2016, SchroederReview2016}, and possibly starting as low as $1\,$PeV \cite{Balagopal2018}.
This advance in technology enables hybrid arrays combining \Xmax\ and muon measurements in the same experiment to maximize the total accuracy for CRs in the PeV-EeV range.

In the next decade, the inclusion of other observables in combination with modern analysis techniques, such as machine learning \cite{Erdmann:2017str,Erdmann:2019nie}, needs to be investigated to increase the efficiency and reconstruction accuracy.
GCR observatories have already played a crucial role in the development and test of new techniques, like radio \cite{2005Natur.435..313F, Ardouin:2009zp, Acounis:2012dg, Schellart:2013bba, TunkaRex_NIM} and microwave \cite{2014PhRvL.113v1101S} measurements. 
Radio measurements also provide a new tool to compare the energy scales of different experiments \cite{Apel:2016gws, Aab:2016eeq, Gottowik:2017wio}. 

For understanding the origin of GCRs, more accurate air-shower instrumentation needs to be accompanied by multi-messenger astronomy of GCRs, photons, and neutrinos. 
First, the detection of photons and neutrinos can directly reveal sources of GCRs~\cite{Bai:2014kba,Anchordoqui:2016dcp}. 
For energies up to about 100~TeV, gamma-ray observatories have already discovered a few sources \cite{2018EPJP..133..324D, FunkReview2015, 2018arXiv180810495P, 2015CRPhy..16..587D, 2015CRPhy..16..610D, 2015arXiv151005675H}. 
PeV photons have not yet been observed, but may be emitted by the Galactic Center if it accelerates the most energetic GCRs~\cite{Aharonian:2006au, Balagopal2018}.
Second, an accurate knowledge of the absolute flux and mass composition of CRs is important for the interpretation of other messengers, e.g., to understand the background for neutrino measurements in IceCube \cite{Gaisser:2016obt}. 
Third, GCR, neutrino, and photon measurements need to provide a consistent picture.
Measurements of diffuse gamma-ray and neutrino fluxes give insight in the propagation of GCR \cite{Gaggero:2015xza, Chantell:1997gs, Apel:2017ocm, Albert:2018vxw, Pandya:2017akd}. 
In particular, the diffuse gamma-ray flux observed by {\it Fermi}-LAT~\cite{Ackermann:2014usa} is slightly in tension with the soft power-law neutrino flux deduced from the IceCube High Energy Starting Events~\cite{Aartsen:2018fqi}, which may indicate a non-negligible Galactic contribution to the astrophysical neutrino flux~\cite{Murase:2015xka, Aartsen:2017ujz, Neronov:2018ibl}.
The observation of high-energy Galactic neutrinos will provide a lower limit on the maximum acceleration energy~\cite{Anchordoqui:2013qsi}, and may directly reveal sources, similar to the observation of Galactic gamma rays.

\section{Experimental Plans for the Next Decade}\vskip-2mm
Multi-messenger astronomy requires a solid foundation in accurate instrumentation for all cosmic messengers. 
While major investments in detectors for high-energy photons and neutrinos are foreseen, the required increase in accuracy for GCRs can be obtained for a comparably low cost.
This improvement is possible by upgrading existing air-shower detectors.
Furthermore, detectors built primarily for photons and neutrinos can be enhanced to provide accurate CR measurements at the same time, transforming them to multi-messenger observatories. 

The scientific questions can be targeted with dedicated efforts for increasing the accuracy in the relevant energy range, in particular enhancements of the IceCube surface array, the Telescope Array and the Pierre Auger Observatory.
Several future observatories with different main objectives will also contribute to GCR science, e.g., the SKA \cite{HuegeSKA_ICRC2015}, TAIGA \cite{TAIGA_2014}, and GRANDproto300 \cite{Alvarez-Muniz:2018bhp}.
Their accuracy for cosmic rays will depend on whether they will be equipped with precise \Xmax\ and muon instrumentation. 
Dense particle-detector arrays at high observation altitude optimized for gamma-astronomy, such as HAWC \cite{Alfaro:2017cwx}, a HAWC-like southern observatory \cite{DuVernois:2015ima}, or LHASSO \cite{He2018} will also have a high precision for GCRs in the PeV region. 

IceCube with its surface array IceTop covers the complete range of high-energy GCRs from below 1~PeV to beyond 1~EeV \cite{IceCube:2012nn}. 
The simultaneous measurement of low-energy particles at the surface and high-energy muons in the ice offers unique opportunities for the study of hadronic interactions \cite{Aartsen:2015nss}, and the search for PeV photons \cite{Aartsen:2012gka}. 
A planned enhancement by a scintillator-radio hybrid array will significantly increase the accuracy and sky coverage of IceTop \cite{ScintillatorsICRC:2017, Schroder:2018dvb, Haungs:2018}. 
Air-Cherenkov detectors can further enhance its accuracy around a few PeV and below \cite{Auffenberg:2017vwc}. 
Finally, a planned expansion of IceCube will increase the exposure by an order of magnitude \cite{Ackermann:2017pja}.

For studying the Galactic-to-extragalactic transition range, the most important contributions in the next years are expected from the low-energy extensions of the two leading observatories for ultra-high-energy CRs: the Pierre Auger Observatory \cite{AugerNIM2015} and the Telescope Array \cite{Abbasi:2018xsn}.
The TALE fluorescence detector at TA covers the energy range from about $2\,$PeV to beyond $1\,$EeV in monocular mode \cite{Abbasi:2018xsn}, and the range above $100\,$PeV in an recently enabled hybrid mode with scintillation detectors \cite{Ogio:2018hyq}.  
To reach energies below $1\,$PeV, another planned addition to TALE is the Non-Imaging Cherenkov Array (NICHE)\cite{Bergman:2017nrx}.
The Auger enhancements reach below $100\,$PeV. 
Their accuracy will be increased by installing underground muon detectors \cite{PierreAugur:2016fvp}, complementing the upgrade of the surface array \cite{Aab:2016vlz}, and the fluorescence \cite{AugerNIM2015}, and radio detectors \cite{Aab:2015vta}. 

The increase in accuracy, exposure, and sky coverage provided by these experiments will bring unprecedented sensitivity for the search of mass-sensitive anisotropies. 
Hence, the contribution of GCRs to multi-messenger astrophysics will be lifted to a new level providing a real chance finally to discover the most energetic accelerators in our Milky Way.

\pagebreak


\bibliographystyle{utphys.bst}
\bibliography{references}

\end{document}